\pdfoutput=1
\documentclass[11pt]{article}
\pdfoutput=1
\usepackage[margin=1in]{geometry}
\usepackage{mathptmx}
\usepackage{titlesec}
\usepackage{etoolbox}
\usepackage{setspace}
\usepackage{graphicx}
\usepackage{amsmath,amssymb,amsfonts}%
\usepackage{amsthm}%
\usepackage{newtxtext}
\usepackage{newtxmath}
\usepackage{hyperref}
\usepackage{float}
\usepackage{soul}
\hypersetup{
	colorlinks = true,
	urlcolor   = blue,
	citecolor  = black,
}

\usepackage{multirow}%
\usepackage{mathrsfs}%
\usepackage[title]{appendix}%
\usepackage{xcolor}%
\usepackage{textcomp}%
\usepackage{manyfoot}%
\usepackage{booktabs}%
\usepackage{algorithm}%
\usepackage{algorithmicx}%
\usepackage{algpseudocode}%
\usepackage{listings}%
\usepackage[authoryear]{natbib}
\setcitestyle{aysep={}}
\usepackage{caption}
\usepackage{subcaption}

\newcommand{\RomanNumeralCaps}[1]
\linenumbers

\title{Maximum Likelihood Particle Tracking in Turbulent Flows via Sparse 
Optimization}
\author{
	Griffin M. 
	Kearney$\mathrm{^{1,2}}$\footnote{gkearneyengineering@gmail.com}, 
	Kasey M. Laurent$\mathrm{^{1}}$\footnote{klaurent@syr.edu}, 
	Makan Fardad$\mathrm{^{1}}$\footnote{makan@syr.edu} \\
	\vspace{0.5cm} \\
	$\mathrm{^1}${\it Syracuse University, Syracuse, NY 13244} \\
	$\mathrm{^2}${\it OpB Data Insights LLC, Syracuse, NY 13224}
}

\begin{document}
	\maketitle

	\vskip\bigskipamount 
	\vskip\medskipamount
	\leaders\vrule width \textwidth\vskip0.4pt 
	\vskip\bigskipamount 
	\vskip\medskipamount
	\nointerlineskip
	\begin{abstract}
		Lagrangian particle tracking is essential for characterizing turbulent 
		flows, but inferring particle acceleration from inherently noisy 
		position data remains a significant challenge. Fluid particles in 
		turbulence experience extreme, intermittent accelerations, resulting in 
		heavy-tailed probability density functions (PDFs) that deviate strongly 
		from Gaussian predictions. Existing filtering techniques, such as 
		Gaussian kernels and penalized B-splines, implicitly assume 
		Gaussian-distributed jerk, thereby penalizing sparse, high-magnitude 
		acceleration changes and artificially suppressing the intermittent 
		tails. In this work, we develop a novel maximum likelihood estimation 
		(MLE) framework that explicitly accounts for this non-Gaussian 
		intermittency. By formulating a modified Gaussian process to model the 
		random incremental forcing, we introduce a sparse optimization scheme 
		utilizing a convex $\ell_1$-relaxation. To overcome the numerical 
		stiffness associated with high-order difference operators, the problem 
		is efficiently solved using an iteratively reweighted least squares 
		(IRLS) algorithm. The proposed filter is evaluated against direct 
		numerical simulation (DNS) data of homogeneous, isotropic turbulence 
		($\mathrm{Re}_{\lambda} \simeq 310$). Results demonstrate that the IRLS 
		approach consistently outperforms state-of-the-art discrete MLE, 
		continuous MLE, and B-spline methods, yielding systematic reductions in 
		root-mean-squared error (RMSE) across position, velocity, and 
		acceleration. Most importantly, the proposed framework succeeds in 
		better recovering the heavy-tailed statistical structure of both 
		acceleration and acceleration differences (jerk) across temporal 
		scales, preserving the physical intermittency characteristic of 
		high-Reynolds-number turbulent flows that baseline methods severely 
		attenuate.
	\end{abstract}
	
%%==================================%%
%%           INTRODUCTION           %%
%%==================================%%
\section{Introduction}\label{sec1}
\label{sec:intro}

Understanding the dynamics of fluid particles in turbulence is essential for 
characterizing mixing, dispersion, and energy transfer across scales. 
Turbulence consists of spatially and temporally localized regions of intense 
velocity gradients, where fluid particles are subject to extreme accelerations 
as they pass through coherent structures such as vortex tubes. These 
interactions result in intermittent acceleration statistics, where probability 
density functions (PDFs) consist of heavy tails that deviate significantly from 
Gaussian predictions. Experimental Lagrangian particle tracking techniques have 
made it possible to capture detailed particle trajectories, but acceleration 
must be inferred from inherently noisy position data, requiring filtering 
methods that inevitably embed assumptions about the nature of the particle 
motion. To accurately estimate accelerations from noisy position data, 
filtering must be informed by the physics of turbulent fluctuations to avoid 
biasing acceleration statistics.

Acceleration PDFs of fluid particles in turbulence exhibit a peak at zero and 
heavy tails, deviating strongly from a Gaussian distribution. \citet{Voth2002} 
reported experimental one-component acceleration PDFs with kurtosis values 
exceeding 40, dependent on the Reynolds number. \citet{Mordant2004} extended 
measurements to probabilities as low as $10^{-7}$ and observed persistent heavy 
tails in the PDF of accelerations, extending beyond 50 standard deviations. The 
existence of heavy tails implies rare, intense acceleration events, due to 
strong intermittency inherent to small-scale turbulence. Extreme accelerations 
are produced when fluid particles encounter intense, small-scale flow 
structures which impart sudden and large forces on the particle 
\citep{bentkamp2019}.

Early efforts to estimate acceleration from noisy particle tracks obtained from 
experiments primarily focused on noise suppression without explicitly 
incorporating flow physics. Early particle tracking studies often used 
low-order polynomial fits over a sliding window of a few points to compute 
velocities and accelerations \citep{Voth2002, ayyalasomayajula2006lagrangian, 
luthi2005lagrangian}. The polynomial order and window length must be carefully 
chosen; if it is too low, the curvature of the trajectory cannot be captured, 
and if it is too high, the filter overfits and results in nonphysical 
oscillations. Another common technique is Gaussian kernel convolution, 
introduced by \citet{Mordant2004}. The need to convolve with a kernel over 
multiple points means that very short particle tracks cannot be filtered. Short 
tracks frequently belong to fast-moving particles, which inherently experience 
larger accelerations \citep{Sawford2003}. Omitting these trajectories leads to 
a selection bias in which extreme events are filtered out, causing a 
suppression of the acceleration PDF tails. This bias can be mitigated by 
oversampling; however, achieving the necessary sample rates (100 
samples/Kolmogorov time scale, or tens to thousands of frames per second 
\citep{Voth2002}) is technologically challenging, especially while maintaining 
a large measurement volume. 

In recent years, more sophisticated smoothing techniques have been developed to 
improve acceleration estimates from noisy position data. Most notable is the 
TrackFit (or FlowFit in the Eulerian case) algorithm, described in 
\citet{Gesemann2018}, which uses penalized B-spline fitting. In this approach, 
each particle trajectory is approximated by a cubic B-spline, whose shape is 
determined by minimizing a cost function that penalizes deviations from data as 
well as excessive curvature. \citet{Gesemann2021} showed that B-spline fitting 
reduces the sensitivity of acceleration statistics to filter length and 
resolves smaller-scale features compared to the Gaussian kernel method. 
\citet{Lawson2018} found that with this B-spline approach, the estimated 
acceleration variance and kurtosis plateau to stable values for a wide range of 
filter parameters whereas the Gaussian kernel approach produced results that 
varied significantly. The TrackFit approach effectively imposes a penalty on 
the third derivative of the trajectory, discouraging large values in jerk. In 
\citet{Kearney2024}, we developed a Bayesian Maximum Likelihood Estimation 
(MLE) framework that incorporates joint process and measurement randomness. The 
method explicitly models the jerk of the fluid particle as a Gaussian random 
process rather than treating it implicitly, and we found the observed 
performance to be comparable to that of the B-splines. In parallel, we 
developed a related MLE framework for general data systems \citep{kearFar2024}, 
that likewise models jerk as a Gaussian process, to construct optimal spline 
functions and denoising conditions which yield continuous-time smooth 
trajectories. The framework uses a control theoretic approach which enables the 
rigorous modeling of both stochastic process dynamics and measurements, making 
it well suited for the fluid particle tracking problem considered in the 
present work.

Observed deviations from Gaussianity have led to the development of stochastic 
models that explicitly incorporate intermittency. Kolmogorov 1941 theory 
assumes homogeneous, isotropic self-similar turbulence with no intermittency 
\citep{kolmogorov1941local}, while the refined similarity hypothesis introduces 
a fluctuating small-scale dissipation rate \citep{kolmogorov1962refinement}. 
Assuming the local dissipation is log-normal, the acceleration PDF becomes a 
mixture of Gaussians, yielding heavy-tailed distributions consistent with 
experiments \citep{Mordant2004}. However, this framework fails to capture 
high-order statistics \citep{frisch1995turbulence}. The Sawford model 
\citep{sawford1991reynolds} provided an early stochastic description of 
velocity and acceleration using a second-order autoregressive process, but its 
Gaussian acceleration assumption prevents reproduction of intermittency. To 
overcome this limitation, a number of other Lagrangian models have been 
developed. \citet{beck2001dynamical} modeled fluctuating acceleration variance 
with a $\chi^2$ distribution, producing power-law tails but a divergent fourth 
moment inconsistent with experiments \citep{Mordant2004}. Later extensions in 
\citet{beck2003} adopted a log-normal variance, producing better agreement with 
observed flatness and tail behavior for acceleration PDFs. Similarly, 
\citet{reynolds2003superstatistical} incorporated a fluctuating dissipation 
variable within a Markovian framework, assuming Gaussian acceleration 
conditioned on local dissipation in accordance with refined similarity theory, 
successfully reproducing heavy tails and Reynolds-number dependence consistent 
with experiments \citep{LaPorta2001, Voth2002}. Motivated by DNS evidence that 
conditional acceleration remains intermittent, 
\citet{lamorgese2007conditionally} introduced a conditionally cubic-Gaussian 
extension that better captures PDF shape and tails. Multifractal cascade 
approaches have also been applied, with \citet{arimitsu2004multifractal} 
deriving acceleration PDFs via generalized entropy maximization and 
\citet{chevillard2006lagrangian} predicting Reynolds-dependent acceleration 
statistics consistent with DNS and experiments. 

In general, the developed models for Lagrangian particle motion in turbulence 
indicate that the jerk will be non-Gaussian, contradicting the assumed Gaussian 
distribution for the TrackFit filter and MLE filters. As a result, occasional 
large spikes in acceleration change will be penalized and under-represented in 
data processed with TrackFit. In this paper, we develop and evaluate an 
improved MLE framework for recovering particle acceleration from noisy position 
data without imposing a Gaussian-distributed jerk. This filtering approach 
allows for greater flexibility in capturing the rapid, intermittent changes in 
acceleration that are characteristic of turbulent flows. By formulating the 
problem in this framework, we can incorporate non-Gaussian models for 
acceleration differences. This approach offers a way to evaluate the role of 
prior assumptions in trajectory filtering and provides a foundation for more 
physically consistent acceleration estimates, improving our ability to analyze 
Lagrangian turbulence from experimental data. 

The current gap lies in the disconnect between physics and filtering practice. 
While particle velocity differences and acceleration in turbulent flows have 
been extensively characterized, the behavior of the acceleration differences 
and jerk has received relatively little attention. Jerk reflects how rapidly 
the forces acting on a fluid particle change over time, revealing how particles 
interact with fine-scale structures in the flow. Heavy tails in PDF of jerk 
imply large, abrupt changes in acceleration (for example, a particle suddenly 
entering or exiting a strong vortex filament). Existing models imply this 
behavior. For the Sawford model \citep{sawford1991reynolds}, jerk is not 
well-defined, due to the white-noise forcing of acceleration. While an explicit 
PDF for jerk is not given for later models in \citet{beck2001dynamical, 
beck2003, reynolds2003superstatistical, lamorgese2007conditionally}, one can 
infer that jerk inherits intermittency, and large jerks are expected when there 
are sharp changes in the local dissipation rate. 
\citet{arimitsu2004multifractal} proposed a multifractal approach grounded in 
generalized entropy maximization. They explicitly derive the acceleration PDF, 
which closely matches experimental data \citep{Mordant2004}. This model tends 
to perform better than other models in reproducing the core of the acceleration 
distribution, whereas log-normal models tend to overestimate. Additionally, the 
approach described in \citet{chevillard2006lagrangian} produces fluid velocity 
that is infinitely differentiable at finite Reynolds numbers; the model 
predicts that jerk will also display intermittent fluctuations.

The remainder of this paper is organized as follows: Section \ref{sec:tech-dev} 
details the mathematical formulation of the maximum likelihood estimation 
framework, introducing the modified Gaussian process model for intermittent 
jerk and the subsequent convex relaxation of the optimization problem. Section 
\ref{sec:methods} outlines the methodology, including the numerical 
implementation of the iteratively reweighted least squares (IRLS) solver and 
the error quantification metrics applied to the synthetic turbulent trajectory 
data. Section \ref{sec:results} presents the results and discussion, comparing 
the proposed IRLS filter's performance against existing state-of-the-art 
filtering techniques. Finally, Section \ref{sec:conclusion} offers concluding 
remarks and highlights avenues for future work, including the potential for 
real-time state estimation and establishing rigorous links between optimization 
hyperparameters and physical flow constants.

%%==================================%%
%%      TECHNICAL DEVELOPMENT       %%
%%==================================%%
\section{Technical Development}
\label{sec:tech-dev}
In this section we will standardize the maximum likelihood analysis presented 
in our past work \citep{Kearney2024} by developing the general problem in terms 
of modern optimization in Section \ref{ssec:opt-frame}.
This treatment provides a sufficiently general framework for application to 
many interesting particle tracking data systems, including those with 
non-Gaussian randomness.

In Section~\ref{ssec:mod-gauss}, we develop a specific realization of the 
general framework by modifying the assumed statistical structure of the process 
dynamics. Here, we challenge the notion that standard Gaussian distributions 
are accurate representations of the random incremental forcing, or jerk, on a 
fluid particle.

The optimization problem derived in Section \ref{ssec:modeling-mle} using the 
nonstandard distribution is nonconvex, and thus it is difficult to analyze in 
its original form.
In Section \ref{ssec:analysis-mle} we apply relaxation techniques to derive 
convex approximations which are easily solvable using standard convex 
optimization techniques.
The solution of the convex relaxation is discussed further when results are 
presented in Section \ref{sec:results}.

\subsection{A Standard Maximum Likelihood Framework}
\label{ssec:opt-frame}
In previous work \citep{Kearney2024}, we developed a maximum likelihood data 
filtering scheme using Bayesian arguments.
However, the approach is better standardized using optimization-theoretic 
treatments in which the framework is stated using optimization objectives and 
constraints.
The objective of the scheme is a measure of the total probability, which 
depends on both the distribution of process randomness and the measurement 
error.
The constraints are used to enforce the dynamics underlying the data system as 
well as the structure of the measurement errors.

We consider particle trajectories that are discrete time series of length $T$, 
where $T$ denotes the total number of samples.
We define $x$ to be the $T$-dimensional vector representing the true trajectory 
of the particle in a single positional dimension, absent measurement error, and 
$y$ to be the $T_m$-dimensional vector representing noisy measurements of $x$. 
Here $T_m$ denotes the number of discrete measurements gathered to perform the 
smoothing, and in direct sampling schemes we have $T_m = T$.
We define $v$ to be a random vector-valued variable of dimension $T_d$. $T_d$ 
denotes the number of discrete points at which the dynamics are computed; for 
instance, when applying standard finite difference equations $T_d < T$ since 
central differencing cannot be applied outright at the edges of the vector $x$.
The vector $v$ represents the process noise and is used to model uncertainty in 
the dynamics. 
Similarly, we define $w$ to be a random vector-valued variable of dimension 
$T_m$ that represents the measurement noise and which is used to model 
uncertainty in observations.
The random processes $v$ and $w$ are the core of the system likelihood as they 
are non-deterministic.
We denote their probability density functions as $\rho_v(v)$ and $\rho_w(w)$, 
respectively.
Furthermore, we assume that $v$ and $w$ are independent random variables so 
that the joint probability density function ($\rho(v,w)$) is the product, given 
by
\begin{equation}
	\label{eqn:joint-pdf}
	\rho(v,w) = \rho_v(v) \rho_w(w).
\end{equation}

The natural logarithm is a monotonic function, so without loss of generality we 
use
\begin{equation}
	\label{eqn:opt-objective}
	J(v,w) = \log \rho(v,w) = \log \rho_v(v) + \log \rho_w(w),
\end{equation}
as the maximization objective of our framework, and we next move to consider 
constraints.
We assume that the dynamics are modeled using a stochastic process of the form
\begin{equation}
	\label{eqn:stoch-process}
	A x = v,
\end{equation}
where $A$ is a matrix of dimension $T_d \times T$; ``the dynamics matrix''.
The model described in (\ref{eqn:stoch-process}) is useful as it can be used to 
describe the discretized form of any linear differential equation.
Similarly, we assume that noisy observations are of the form
\begin{equation}
	\label{eqn:stoch-measurements}
	y = B x + w,
\end{equation}
where $B$ is a matrix of dimension $T_m \times T$; ``the observation matrix''.
Equations (\ref{eqn:stoch-process}) and (\ref{eqn:stoch-measurements}) comprise 
the constraints of the optimization problem, enforcing the problem dynamics and 
measurement system in the solution.
The general problem in our framework is written
\begin{align}
	\label{eqn:gen-opt-problem}
	\underset{x,v,w}{\max} & \hspace{0.25cm} J(v,w)\\
	\notag
	\mathrm{s.t.} & \hspace{0.25cm} A x = v\\
	\notag
	& \hspace{0.25cm} y = Bx + w.
\end{align}

The framework provided in (\ref{eqn:gen-opt-problem}) is populated by 
specifying the dynamics matrix $A$, observation matrix $B$, and the probability 
distributions $\rho_v$ and $\rho_w$ of $v$ and $w$, respectively.
We will consider directly measured systems where $B = I$, the $T \times T$ 
identity matrix, in the remainder of this work. 
In the next section we examine the TURB-Lagr data to demonstrate that $v$ 
exhibits a nonstandard distribution.
The distribution takes the form of a modified Gaussian distribution, one which 
is described by a Gaussian distribution with additional atomic probability 
weight at the point $v = 0$.
The hybrid discrete-continuous nature of this type of modified distribution 
necessitates additional considerations when incorporating it in our 
optimization framework as it does not permit a continuous probability density 
function.

%%%%%%%%%%%%%%%%%%%%%%%%%%%%%%%%%%%%%%%%%%%%%%%%
%MODIFIED GAUSSIANS
%%%%%%%%%%%%%%%%%%%%%%%%%%%%%%%%%%%%%%%%%%%%%%%%
\subsection{Modified Gaussian Distribution Modeling in Turbulent Data}
\label{ssec:mod-gauss}
In (\ref{eqn:gen-opt-problem}) we prescribe distributions that govern the 
process randomness ($v$) and the measurement error ($w$) which ultimately 
determine the structure of the objective $J$.
A well-established approach, and one that we utilized in our previous work, is 
to assume that both sources of uncertainty were Gaussian in nature, which leads 
to a quadratic program.
However, despite our imposing these processes within a third order system (e.g. 
Gaussian jerk), we observed that this leads to muted acceleration distributions 
in smoothed trajectories; a persistent problem among existing methods.
Our prior approach showed improved performance over standard convolution 
smoothing as developed in \citep{Mordant2001} and performed comparably, if not 
slightly better, than B-spline type methods \citep{Gesemann2018}. Importantly, 
our method was built upon physical first principles that we have standardized 
in the prior section. The comparable performance in B-spline methods rely on 
empirical motivations. We emphasized in our initial work that our framework 
enables us to incorporate general modifications in a way that other methods 
cannot. We proceed to do so here by developing a \emph{modified Gaussian 
process} (MGP) smoother using our approach.

We maintain the standard Gaussian model for the measurement noise $w$ as 
before, but consider a modified variant in lieu of a standard Gaussian to model 
$v$.  We again utilize a third order system in which the equation $A x = v$ is 
prescribed as a system of finite differences that describe jerk estimates which 
will be elaborated upon in Section \ref{ssec:modeling-mle}.
The MGP is developed conditionally, to describe the treatment let us begin by 
assuming $v_k$ is a random \emph{modified Gaussian variable} (MGV) 
corresponding to such a process and which constitutes the $k$-th component of 
the vector $v$. 
The variable $v_k$ is assumed to take the value $0$ with probability $p_0 \in 
[0,1]$ and otherwise it takes a value governed by the distribution 
$\mathrm{N}(0,\sigma_v^2)$. For convenience we denote this modified 
distribution as $\tilde{\mathrm{N}}(p_0, 0, \sigma_v^2)$.

Under this definition, the system dynamics can be viewed as ``stubborn 
Gaussian'': it has some base likelihood $p_0$ in which nothing happens (e.g. $v 
= 0$), inducing intermittency in the dynamics. When a non-zero jerk is present 
in the system (with likelihood $1-p_0$), then the non-zero value is taken from 
a Gaussian distribution. In essence we are adding extra weight to the $0$ 
likelihood and this in turn enables us to increase $\sigma_v$ to accommodate 
increased volatility when force-changing events occur.  We treat $v$ as a 
vector with identical independently distributed (IID) components, each of which 
is a MGV.

%---------------------------------------------------------------------
%Modeling MLE
%---------------------------------------------------------------------
\subsection{Modeling of the Maximum Likelihood Estimator}
\label{ssec:modeling-mle}
Modeling the maximum likelihood estimator boils down to quantifying the 
likelihood of the observation of such a vector $v$.
We utilize a simple central difference approximation of a third order system in 
(\ref{eqn:stoch-process}) using the matrix
\begin{equation}
	\label{eqn:A-def}
	A = \frac{1}{\Delta t^3}
	\begin{bmatrix}
		-1 &  3 & -3 &  1 &  0 & \dots & 0 \\
		0 & -1 &  3 & -3 &  1 & \dots & 0 \\
		\vdots & \ddots & \ddots & \ddots & \ddots & \ddots & \vdots \\
		0 & \dots &  0 & -1 &  3 & -3 &  1
	\end{bmatrix}
\end{equation}
where $\Delta t$ denotes the sampling period.
This matrix is of dimension $T-3 \times T$ which requires that $v$ is of 
dimension $T-3$.
Consider $v$ with $0 \leq n \leq T-3$ non-zero components, corresponding to a 
non-zero index set $ \mathscr{N} \subseteq \{1,2, ..., T-3\}$; or in other 
words $|\mathscr{N}| = n$. 
Under the IID assumption the $n$ nontrivial components are governed by the 
distribution
\begin{equation}
	\label{eqn:MGV-given-n-dist}
	\rho_v(v \mid \mathscr{N}) = \Big(\frac{1}{\sqrt{2 \pi} \sigma_v}\Big)^{n} 
	e^{-\frac{||v||^2}{2 \sigma_v^2}},
\end{equation}
where the full norm of the vector $v$ is used in the exponent since the norm of 
the full vector is equal to the norm of its non-zero components, trivially.
The probability of this non-zero index set is given by
\begin{equation}
	\label{eqn:N-dist}
	p(\mathscr{N}) = \Big(p_0 \Big)^{T-3-n} \Big( 1 - p_0 \Big)^{n} = \Big( p_0 
	\Big)^{T-3} \Big( \frac{1 - p_0}{p_0} \Big)^n.
\end{equation}
Strictly speaking, it is possible for $\mathrm{card}(v)$ (the cardinality of 
$v$) to be less than $n$; the distribution $\mathrm{N}(0,\sigma_v^2)$ has 
support at the value $0$. We treat this as pathological, for it occurs with 
probability zero. We therefore assume $n = \mathrm{card}(v)$ which enables us 
to write the joint distribution
\begin{equation}
	\label{eqn:v-joint-dist}
	\rho_v(v \cap \mathscr{N} ) = p(\mathscr{N}) \rho_v( v \mid \mathscr{N}) = 
	(p_0)^{T-3} \Big( \frac{1 - p_0}{\sqrt{2 \pi} \sigma_v p_0} 
	\Big)^{\mathrm{card}(v)} e^{-\frac{||v||^2}{2\sigma_v^2}}.
\end{equation}

As in our prior work, we assume that the measurement errors are IID Gaussian 
random variables with zero mean and variance $\sigma_w^2$.
This yields the distribution
\begin{equation}
	\label{eqn:w-dist}
	\rho_w(w) = \Big( \frac{1}{\sqrt{2 \pi} \sigma_w} \Big)^T 
	e^{-\frac{||w||^2}{2 \sigma_w^2}}.
\end{equation}
We assume that the process and measurement randomness are independent so that 
the joint distribution is given by the product of (\ref{eqn:v-joint-dist}) and 
(\ref{eqn:w-dist}).
We use this product, and after discarding immaterial constant factors, it 
yields the objective $J(v,w)$
\begin{equation}
	\label{eqn:J-MLE}
	J(v,w) = -\frac{\| v \|^2}{2 \sigma_v^2} - \frac{\| w \|^2}{2 \sigma_w^2} - 
	\mathrm{log}\Big( \frac{\sqrt{2 \pi} \sigma_v p_0}{1-p_0}\Big) 
	\mathrm{card}(v).
\end{equation}
Finally, after absorbing a negative sign to exchange the max for a min, this 
yields the maximum likelihood estimator given by the optimization framework:
\begin{align}
	\label{eqn:mle-main}
	\underset{x,v,w}{\min} & \hspace{0.15cm} \frac{\| v \|^2}{2 \sigma_v^2} + 
	\frac{\| w \|^2}{2 \sigma_w^2} + \mathrm{log}\Big( \frac{\sqrt{2 \pi} 
	\sigma_v p_0}{1-p_0}\Big) \mathrm{card}(v)\\
	\notag
	\mathrm{s.t.} & \hspace{0.15cm} A x = v\\
	\notag
	& \hspace{0.15cm} y = x + w.
\end{align}
Problem (\ref{eqn:mle-main}) is not convex in general due to the cardinality 
term of the objective.
Solving this system is therefore non-trivial but convex relaxations do exist 
which enable efficient computation of practically relevant solutions.

A key parameter within our new problem is the cardinality coefficient, given by
\begin{equation}
	\label{eqn:gamma-def}
	\gamma = \mathrm{log}\Big( \frac{\sqrt{2 \pi} \sigma_v p_0}{1-p_0}\Big).
\end{equation}
If $\gamma < 0$ then the cardinality no longer induces a penalty and instead 
rewards non-zero components in $v$. When this is the case, a standard quadratic 
programming solver (convex) is applied since the cardinality term can be 
omitted; this is equivalent to the discrete MLE solver we developed in our 
prior work.
We focus on the more interesting case when $\gamma > 0$, which is equivalent to 
the condition
\begin{equation}
	\label{eqn:nonconv-cond}
	\frac{\sqrt{2 \pi} \sigma_v p_0}{1-p_0} > 1.
\end{equation}
Equation (\ref{eqn:nonconv-cond}) defines a relationship between $\sigma_v$ and 
$p_0$ that determines when the trade-off between volatility in the non-zero 
stochastic dynamics and the ``stubborness'' of the modified process becomes 
meaningful.
In the next section we will describe how the $1$-norm can be used to relax the 
nonconvex optimization problem (\ref{eqn:mle-main}) when 
(\ref{eqn:nonconv-cond}) is satisfied.

\subsection{Finding Solutions of the Maximum Likelihood Estimator}
\label{ssec:analysis-mle}

Were it not for the existence of the cardinality function in its objective (or 
equivalently, were $\gamma \leq 0$) the optimization problem
\begin{align}
	\label{eqn:l0-problem}
	\underset{x,v,w}{\min} & \hspace{0.15cm} \frac{1}{2 \sigma_v^2} \| v \|^2 + 
	\frac{1}{2 \sigma_w^2} \| w \|^2 + \gamma \mathrm{card}(v)\\
	\notag
	\mathrm{s.t.} & \hspace{0.15cm} A x = v\\
	\notag
	& \hspace{0.15cm} y = x + w.
\end{align}
would be a convex quadratic program, which can be solved in closed-form. The 
inclusion of $\mathrm{card}(v)$ in (\ref{eqn:l0-problem}) penalizes the number 
of non-zero entries in $v$ and, as $\gamma$ increases, renders evermore sparse 
solutions for $v$. However, it also renders the optimization problem 
combinatorial in nature and intractable in general.

It is well-known \citep{BoydVandenberghe2004} that the $1$-norm (also referred 
to as the $\ell_1$-norm) serves as a convex relaxation of 
$\mathrm{card}(\cdot)$ while preserving much of its sparsity-promoting 
properties. We thus consider the optimization problem
\begin{align}
	\label{eqn:l1-problem}
	\underset{x,v,w}{\min} & \hspace{0.15cm} \frac{1}{2 \sigma_v^2} \| v \|^2 + 
	\frac{1}
	{2 \sigma_w^2} \| w \|^2 + \gamma \| v \|_1\\
	\notag
	\mathrm{s.t.} & \hspace{0.15cm} A x = v\\
	\notag
	& \hspace{0.15cm} y = x + w.
\end{align}
which is convex and therefore can be solved efficiently (numerically) for very 
large problems.
The optimization problem posed in (\ref{eqn:l1-problem}) belongs to a class of 
$\ell_1$-regularized least squares problems, widely known in signal processing 
as basis pursuit denoising \citep{chen1998} and in statistics as lasso 
\citep{tibshirani1996}. While the problem is convex, the non-differentiability 
of the $\ell_1$-norm at the origin precludes the use of standard gradient 
descent methods. Two families of iterative algorithms are typically employed to 
solve such large-scale non-smooth problems: splitting methods and reweighted 
least squares.

Splitting methods, most notably the Alternating Direction Method of Multipliers 
(ADMM), introduce auxiliary variables to decouple the non-smooth sparsity 
penalty from the smooth data fidelity term \citep{boyd2011distributed}. This 
separation allows for efficient closed-form updates of each component. However, 
in the context of Lagrangian particle tracking, the dynamic matrix $A$ 
represents a high-order difference operator divided by a small time step 
$\Delta t$. This structure frequently results in systems with high condition 
numbers leading to numerical stiffness. In such stiff regimes, first-order 
splitting methods like ADMM often exhibit slow convergence tails or stalling 
behavior, failing to resolve the high-amplitude, sparse events that 
characterize the heavy tails of the probability distribution.

An alternative approach is the Iteratively Reweighted Least Squares (IRLS) 
algorithm. IRLS handles the non-smooth objective by approximating the 
$\ell_1$-norm with a weighted $\ell_2$-norm, transforming the problem into a 
sequence of smooth, quadratic minimization sub-problems. The primary advantage 
of this formulation is that each sub-problem is a linear system that can be 
solved using direct matrix factorization methods (e.g., Cholesky or LDL 
decomposition). Unlike first-order methods, direct solvers are inherently 
robust to ill-conditioning, ensuring high-precision solutions even when the 
dynamic constraints are numerically stiff. Given our requirement to accurately 
recover extreme acceleration events without smoothing bias, we adopt the IRLS 
framework for the remainder of this work.
%%==================================%%
%%             METHODS              %%
%%==================================%%
\section{Methods}
\label{sec:methods}

% \begin{figure}
	% \centering
	%     \includegraphics[width=0.8\linewidth]{Figs/fig1_da.eps}
	%     \caption{Acceleration differences from DNS}
	%     \label{fig:da_true}
	% \end{figure}

We evaluated the proposed filter on 6,000 simulated three-dimensional fluid 
particle trajectories, each consisting of 1,800 time steps, obtained from a 
direct numerical simulation (DNS) of homogeneous, isotropic turbulence with 
$\mathrm{Re}_\lambda\simeq310$. The simulated trajectories were drawn from the 
TURB-Lagr database \citep{biferale2024turb}, and include particle position, 
velocity, acceleration, and local velocity gradient information. The time step 
between stored samples is $\Delta t\approx\tau_\eta/153$.

Synthetic position noise was added to DNS particle trajectories prior to 
filtering.
We used each of the simulated particle trajectories as a ground truth ($x$) and 
applied simulated zero-mean Gaussian position errors ($w$) to generate 
``noisy'' measurements ($y$).
For each trajectory we apply the smoothing scheme described by 
(\ref{eqn:mle-main}) using the convex relaxation presented in 
(\ref{eqn:l1-problem}).
We simulated the measurement error as white noise with standard deviation 
$\sigma_w = 0.0063$ for all experiments.
The smoothing scheme and simulation procedure operate on each spatial component 
of the particle trajectory independently.

\subsection{Solver Implementation}
\label{ssec:solver-implement}

The optimization problem presented in (\ref{eqn:l1-problem}) involves a 
non-smooth $\ell_1$-regularization term and is characterized by significant 
numerical stiffness. This stiffness arises from the large disparity between the 
weights of the data fidelity term (governed by the small measurement error 
$\sigma_w$) and the sparsity inducing term. While first-order splitting 
methods, such as the Alternating Direction Method of Multipliers (ADMM), are 
commonly used for such problems, they frequently exhibit slow convergence or 
stalling behavior in stiff parameter regimes, potentially smoothing out the 
extreme events (heavy tails) of interest.

To address these challenges, we employ an iteratively reweighted least squares 
(IRLS) algorithm \citep{Daubechies2010}. This approach transforms the 
non-smooth optimization problem into a sequence of smooth, weighted 
least-squares sub-problems. Unlike splitting methods, IRLS allows for the use 
of direct linear solvers, which are robust to ill-conditioning and ensure 
high-precision convergence.

At each iteration $k$, we approximate the objective function in 
(\ref{eqn:l1-problem}) by a quadratic form. The non-smooth penalty $\gamma 
\|v\|_1$ is replaced by a weighted $\ell_2$-norm term $\frac{1}{2} v^T 
\Omega^{(k)} v$, where $\Omega^{(k)}$ is a diagonal weighting matrix updated 
based on the solution from the previous iteration. To consistently approximate 
the combined $\ell_2$ and $\ell_1$ penalties on the jerk $v$, the diagonal 
entries are defined as:
\begin{equation}
	\label{eqn:irls-weights}
	\Omega_{ii}^{(k)} = \frac{1}{\sigma_v^2} + \frac{2\gamma}{|v_i^{(k-1)}| + 
	\epsilon},
\end{equation}
where $v_i^{(k-1)}$ is the $i$-th component of the jerk estimated at iteration 
$k-1$, and $\epsilon$ is a small regularization constant (set to $10^{-6}$) 
effectively acting as a smoothing parameter to prevent numerical instability 
when $v_i$ approaches zero.

Substituting this approximation into the objective, the update step for the 
trajectory $x$ at iteration $k$ becomes the solution to a weighted least 
squares problem:
\begin{equation}
	\label{eqn:irls-subprob}
	x^{(k)} = \underset{x}{\text{argmin}} \left( \frac{1}{2\sigma_w^2} \|y - 
	x\|_2^2 + \frac{1}{2} (Ax)^T \Omega^{(k)} (Ax) \right).
\end{equation}
The optimality condition for (\ref{eqn:irls-subprob}) yields the following 
linear system:
\begin{equation}
	\label{eqn:irls-linear}
	\left( \frac{1}{\sigma_w^2} I + A^T \Omega^{(k)} A \right) x^{(k)} = 
	\frac{1}{\sigma_w^2} y.
\end{equation}
Since $A$ represents a finite difference operator, the matrix $A^T \Omega^{(k)} 
A$ retains a sparse, banded structure. This structure permits the system to be 
solved efficiently using direct sparse Cholesky factorization with a 
computational complexity of $\mathcal{O}(T)$. The algorithm iterates between 
updating the weights via (\ref{eqn:irls-weights}) and solving the system 
(\ref{eqn:irls-linear}) until convergence. This direct solver approach avoids 
the convergence issues of iterative splitting methods in stiff regimes, 
allowing for the accurate recovery of the heavy-tailed acceleration 
distributions characteristic of high-Reynolds-number turbulence.

\subsection{Error Quantification}
\label{ssec:eval-performance}
\label{ssec:error-dists}

The TURB-Lagr simulation data provide ground-truth trajectory positions, 
velocities, and accelerations, enabling direct quantitative evaluation.
We use $x$ to denote the true (one-dimensional) discrete position, $u$ for the 
true discrete velocity, and $a$ for the true discrete acceleration over the 
time interval.
Quantities marked with a hat denote filter estimates obtained after operating 
on noisy measurements. The proposed filters operate component-wise and the 
resulting three data streams are collected to form the discrete time vectors 
$\hat{x}$, $\hat{u}$, and $\hat{a}$, respectively.
These filtered discrete-time vector quantities are used to compute the error 
$e(\cdot)$. For a given trajectory, the instantaneous reconstruction error is 
defined at each discrete time $t$ via:
\begin{equation}
	\label{eqn:error-def}
	e_{\hat{x}}(t) = \hat{x}(t) - x(t),
\end{equation}
where (\ref{eqn:error-def}) is defined on the reconstructed position $\hat{x}$, 
but is used analogously for velocity and acceleration.

To quantify the overall reconstruction accuracy for an individual trajectory, 
we compute the root-mean-squared error (RMSE), 
\begin{equation}
	\label{eqn:RMSE-error}
	\mathrm{E}(\hat{x}) = \sqrt{\frac{1}{T} \sum_{t=1}^T \|e_{\hat{x}}(t)\|^2},
\end{equation}
where again (\ref{eqn:RMSE-error}) is illustrated on the position, but is 
applied to the velocity and acceleration identically. 

Section \ref{ssec:filter-tuning} evaluates the distributions of the position, 
velocity, and acceleration RMSE across a set of candidate filtering schemes. 
Because the IRLS filter depends on two tunable parameters, the analysis also 
quantifies parameter sensitivity and establishes a practical tuning strategy 
that does not rely on knowledge of the true trajectory.

%%==================================%%
%%       EXPERIMENTAL RESULTS       %%
%%==================================%%
\section{Results and Discussion}
\label{sec:results}
We demonstrate the improved performance of the proposed IRLS filter through a 
comparative study with other state-of-the-art methods.
In our preliminary work \citep{Kearney2024}, we demonstrated that the discrete 
MLE filter without intermittency considerably outperformed baseline Gaussian 
smoothing \citep{Mordant2001}. However, our standard Gaussian MLE smoother 
performance was comparable to the B-spline approach \citep{Gesemann2018}.
Accordingly, we use the standard Gaussian MLE filter, B-spline smoothing and an 
additional MLE-based spline (referred to as the continuous MLE filter) that 
uses the general MLE spline construction approach \citep{kearFar2024}.
The MLE-based splines are constructed using standard Gaussian process noise and 
a third-order (i.e. jerk) continuous system consistent with the model used in 
the MLE filter and the proposed method. For all filtering methods, we 
determined the filter parameters through an exhaustive search, seeking to 
minimize the average RMSE.

Critical to this study, we evaluate the recovery of turbulent flow physics by 
examining the statistics of the Lagrangian acceleration and acceleration 
differences. Since the filter operates on position, acceleration is derived via 
second-order central finite differences of the smoothed position estimates. 
We compute the probability density functions (PDFs) of the acceleration 
components, normalized by their respective standard deviations. These 
distributions are plotted on semi-logarithmic axes to explicitly inspect the 
behavior of the tails. The primary criterion for success of the IRLS filter is 
its ability to better reproduce the heavy, non-Gaussian tails observed in the 
DNS ground truth. This is an artifact of the sparse high-jerk events, rather 
than the parabolic shape characteristic of Gaussian smoothing artifacts.

%We next examine the statistics of the reconstructed acceleration signal. We 
%evaluate the acceleration and acceleration differences via the probability 
%density functions (PDFs). We additionally consider the flatness of the 
%acceleration differences... \kml{finish}
\begin{figure}[t!]
	\centering
	\includegraphics{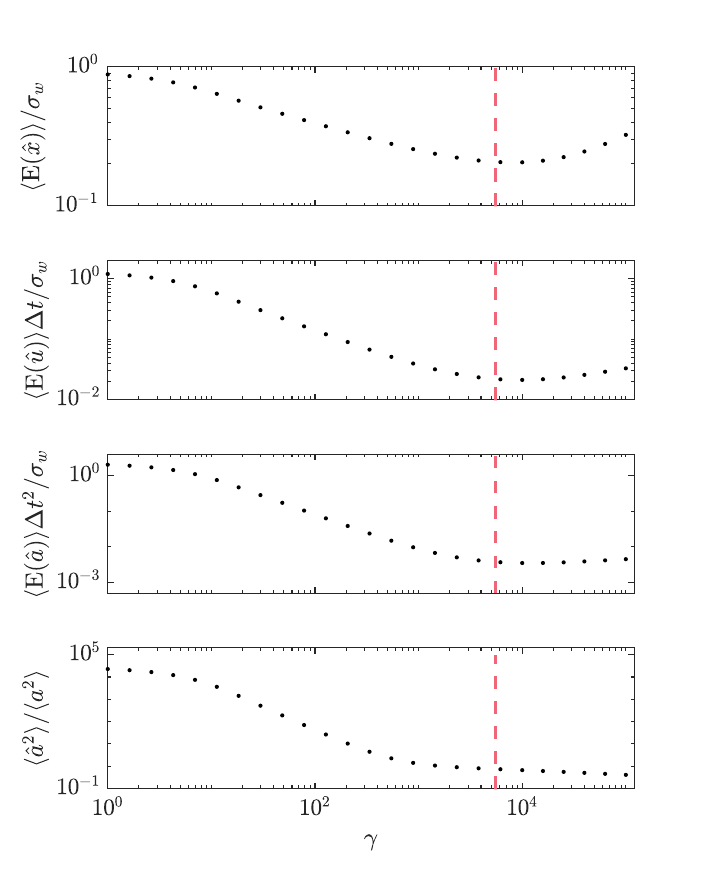}
	\caption{Normalized RMSE from IRLS-filtered trajectories as a function of 
	sparsity penalty parameter $\gamma$ for position (first - top), velocity 
	(second), and acceleration (third) in comparison with reconstructed 
	acceleration statistics as a function of $\gamma$ (bottom). All graphs are 
	plotted on log-log scale and the tuned $\gamma$ value is indicated by the 
	dash pink line. In all RMSE graphs the minima occur shortly after the 
	exponential decay (linear in log-log scale) transition of the acceleration 
	statistics demonstrating that tuning the IRLS filter is achievable even in 
	the absence of ground truth data.}
	\label{fig:gammaSweep}
\end{figure}

\subsection{Filter Tuning and Sensitivity}
\label{ssec:filter-tuning}

The efficacy of the proposed IRLS filter relies on the selection of two 
hyperparameters: the process noise standard deviation $\sigma_v$, which governs 
the assumed volatility of the jerk, and the sparsity penalty parameter $\gamma$ 
as defined in (\ref{eqn:gamma-def}), which controls the strength of the 
$\ell_1$ regularization.
The measurement noise level $\sigma_w$ was held fixed throughout all 
experiments to be representative of a simulated experimental setup.

To determine an optimal tuning configuration, we employed a hierarchical 
grid-search strategy over the parameter space $(\sigma_v, \gamma)$ using a 
representative subset of $200$ trajectories for computational efficiency. We 
evaluated the filter performance using a dual-objective criterion in which we 
quantified both position and acceleration errors of the smoothed trajectory 
versus the known true trajectories taken directly from the TURB-Lagr dataset.
%The resulting error landscapes are presented in Figure 
%\ref{fig:tuning_combined}. The top panels (a) display the initial coarse grid 
%search, which reveals a distinct sensitivity hierarchy. The error gradients 
%are 
%primarily aligned with the $\gamma$ axis, resulting in vertical stratification 
%of the error surface. 
Across wide variations in $\sigma_v$, the resulting error landscapes exhibited 
minimal variation in comparison to the strong gradients observed along the 
$\gamma$ direction.  
%\kml{Kasey: perhaps we can move the heatmaps to supplementary materials or an 
%appendix?} \gmk{Griffin: Yes, let's do this - I added some generic language 
%below to indicate this.} 
In other words, the sparsity penalty almost entirely governs the reconstruction 
quality, whereas the assumed process noise width contributes only marginally. 
We provide the tuning surfaces developed in the process of conducting this 
study in Appendix~\ref{sec:errorMap}.

Guided by this sensitivity hierarchy, we adopted an iterative tuning strategy. 
From our grid-search, we first identified a reasonably small $\sigma_v$, to 
improve numerical stability, in which reconstruction error changed negligibly. 
We then varied $\gamma$ at a higher resolution. The first three panels of 
Figure~\ref{fig:gammaSweep} show the resulting dependence of position, 
velocity, and acceleration average RMSE on $\gamma$. From this, we identified 
an appealing design point at $\sigma_v = 2.2$ and $\gamma = 5528.7$, shown as 
the dashed pink line in Figure~\ref{fig:gammaSweep}.

In practical applications, the true trajectory is not available, and therefore 
RMSE-based optimization is infeasible. To identify a practical tuning approach, 
we examined how the reconstructed acceleration standard deviation varies with 
$\gamma$. We show this trend in the final panel of Figure~\ref{fig:gammaSweep}. 
Examination of Figure~\ref{fig:gammaSweep} (bottom) shows that the acceleration 
statistics decay exponentially (i.e. linear in the plotted log-space) once 
$\gamma$ becomes sufficiently large. Moreover, we note that the minima in the 
RMSE curves all occur shortly after the transition to exponential decay. 
Therefore, in the absence of ground truth data, the filter could be tuned by 
first sweeping over $\gamma$ to identify the region in which the acceleration 
statistics exhibit exponential decay. Selecting $\gamma$ near the onset of this 
region yields near-optimal performance.

Finally, it is tempting to link the optimized penalty parameter back to the 
physical intermittency model, we solve (\ref{eqn:gamma-def}) for the sparsity 
probability $p_0$. This yields the analytical mapping:
\begin{equation}
	\label{eqn:p0-solved}
	p_0 = \frac{e^{\gamma}}{e^{\gamma} + \sqrt{2\pi}\sigma_v} \approx 1 - 
	\sqrt{2 \pi} \sigma_v e^{-\gamma}.
\end{equation}
Using the tuned values of $\gamma = 5528.7$ and $\sigma_v = 2.2$, this 
relationship allows us to quantify the implied probability of jerk 
intermittency, providing a physical interpretation of the flow's intermittency 
captured by the filter. With such a large $\gamma$, the exponential dependency 
exhibited in (\ref{eqn:p0-solved}) yields $p_0$ effectively equal to $1$. This 
result is likely non-physical with the value dominated by the relaxation of the 
$\ell_0$ non-convex problem to the $\ell_1$ convex reweighting scheme.
This is noted because the derived $\sigma_v$ and $p_0$ can be corrupted by the 
nature of the convex relaxation. Relaxation from an $\ell_0$-norm to a 
reweighted $\ell_1$-norm materially affects the underlying model. This 
relaxation enables tractable solution of the optimization problem, and indeed 
we will show in Section \ref{sec:results} that the results produce a strict 
improvement over existing state-of-the-art methods. 
However, in the non-convex problem, the magnitude of non-zero components would 
be penalized according to a function of the form
\begin{equation}
	\label{eqn:penalty-true}
	\frac{||v||_2^2}{2 \sigma_v^2} + \gamma \mathrm{card}(v),
\end{equation}
where the component magnitudes face relatively little penalty \emph{after} they 
are determined to be non-zero.
On the other hand, the relaxation penalizes components according to
\begin{equation}
	\label{eqn:penalty-relaxed}
	\frac{||v||_2^2}{2 \sigma_v^2} + \gamma ||v||_1.
\end{equation}
In the case of (\ref{eqn:penalty-relaxed}), the parameter $\gamma$ imparts a 
heavy penalty on the absolute magnitudes of $v$ which dwarfs the quadratic 
penalty and is suspected to be the root cause of the insensitivity of the 
performance to $\sigma_v$ in the relaxed scheme; as $\sigma_v$ increases the 
quadratic term is further reduced.
% \mf{Makan: Folks actually use ``reweighted $\ell_1$'' to approximate the 
%$\ell_0$ norm -- probably will not work well here because of the numerical 
%issues we have. I think that given our going from $\ell_0$ to $\ell_1$ to 
%(reweighted) $\ell_2$, there is really little to be gained from computing 
%$p_0$.}

% \gmk{Griffin: I agree. Backing out $p_0$ from the very large $\gamma$ is so 
%astronomically close to $1$ that the series of relaxations must have a acute 
%impact on the perceived parameters. }

%\kml{Perhaps we can mention something about in-practice tuning (when the true 
%distribution is unknown) since that was a comment from a referee for our last 
%paper. Would it be worth it to put a figure here that shows the assumed jerk 
%distribution from all the optimally tuned filters vs the true jerk 
%distribution 
%(or maybe in results)?}

%\gmk{Since the compute time is now approx. 30 minutes for the 6000 
%trajectories we can run an illustrative tuning study.  I think that's a good 
%idea. Let's pin down specifics of how we want to show this and then I can 
%curate the analysis / result generation for it.}

\begin{figure}[t]
	\centering
	\includegraphics{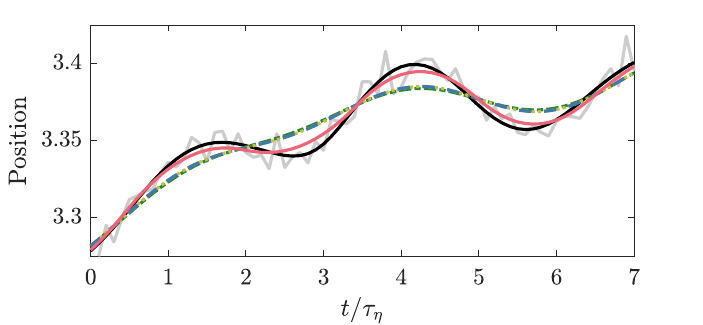}
	\caption{Example of a trajectory. Noisy position data (solid grey) is fed 
	into the IRLS filter (solid pink), the discrete MLE filter (dashed blue), 
	continuous MLE filter (dot-dashed green), and B-splines filter (dotted 
	yellow). Comparing to the
		DNS data (solid black), the IRLS data significantly outperforms the 
		other filters, particularly in regions of extreme accelerations.}
	\label{fig:pos}
\end{figure}

\begin{figure}[t]
	\centering
	\includegraphics{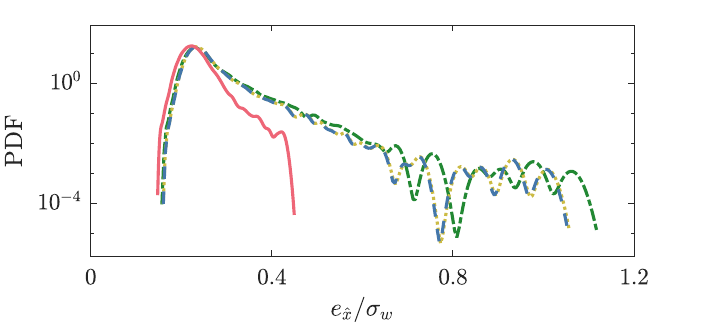}
	\caption{Error distributions for position, normalized by the standard 
	deviation of the measurement noise $\sigma_w$. In all cases, the IRLS 
	method outperforms the discrete MLE, continuous MLE, and B-splines 
	approaches. Coloring and line styles match those used in 
	Figure~\ref{fig:pos}.}
	\label{fig:rmse}
\end{figure}

\begin{figure}[t]
	\centering
	\includegraphics{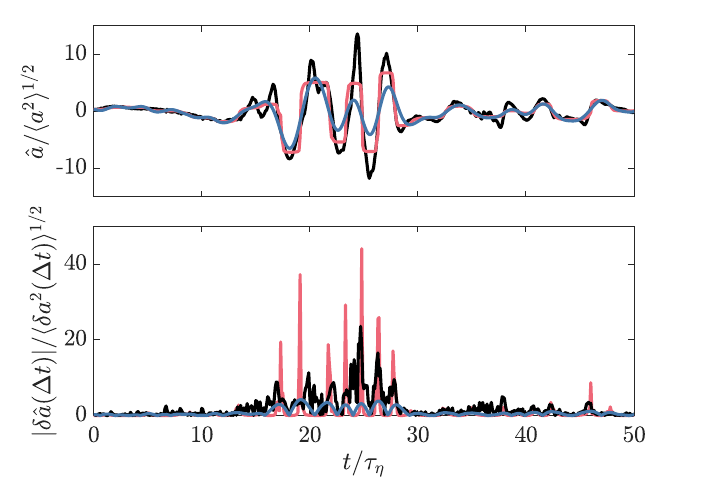}
	\caption{Acceleration time series from a representative trajectory (top) 
	and the corresponding acceleration differences (bottom). Differences are 
	evaluated at the sampling interval $\Delta t$ and serve as a discrete proxy 
	for jerk. All quantities are normalized by the standard deviation of the 
	corresponding quantity from the DNS. Original DNS data is shown in black, 
	the discrete MLE reconstruction in blue, and newly-developed IRLS 
	reconstruction in pink.}
	\label{fig:traj1}
\end{figure}

\subsection{Trajectory Error and Reconstruction}
\label{ssec:illus-app}

Figure~\ref{fig:pos} illustrates a representative segment of a reconstructed 
particle trajectory. As shown, the underlying DNS data (solid black line) 
exhibits rapid, high-curvature transitions indicative of extreme acceleration 
events. The baseline filtering approaches tend to over-smooth these dynamic 
regions, failing to capture the full amplitude of the local spatial 
fluctuations. In contrast, the proposed IRLS filter (solid pink line) tracks 
the ground-truth DNS trajectory with substantially higher fidelity, 
particularly during sharp deviations. By relaxing the assumed Gaussian 
constraint on the jerk and instead penalizing via the $\ell_1$-norm, the IRLS 
approach permits the sparse, high-magnitude changes necessary to resolve these 
intense, intermittent events without overfitting to the high-frequency 
measurement noise (solid grey line).

Figure~\ref{fig:rmse} shows the RMSE distributions for the proposed IRLS filter 
compared to discrete MLE, continuous MLE, and B-splines approaches. All filters 
were optimized to minimize their respective mean RMSE. Across all state 
variables, the IRLS method produces systematic reductions in error. Relative to 
all other filtering approaches, IRLS reduces RMSE by at least 9\% for position, 
15\% for velocity, and 8\% for acceleration. 

We attribute these reductions to the improved ability of the IRLS filter to 
capture rapid temporal fluctuations and intermittent acceleration dynamics. 
Figure~\ref{fig:traj1} (top) shows a representative trajectory characterized by 
large temporal fluctuations in acceleration. Compared with the discrete MLE 
filter from our previous work, the IRLS reconstruction more closely follows the 
high-magnitude oscillations of the DNS signal, which is noticeably attenuated 
by the discrete MLE filter. 

The distinction between the filtering methods becomes more pronounced when 
examining the acceleration differences, defined as $\delta 
\hat{a}(\tau)=\hat{a}(t+\tau)-\hat{a}(t)$, and shown in Figure~\ref{fig:traj1} 
(bottom). Evaluated at the sampling interval $\tau = \Delta t$, these 
differences serve as a discrete proxy for jerk. The discrete MLE filter yields 
small and smoothly varying acceleration differences, indicating that rapid 
transitions in acceleration are strongly suppressed. In contrast, the IRLS 
filter produces intermittent, high-magnitude variations in $\delta \hat{a}$, 
that more closely resemble the DNS signal. Although the IRLS reconstruction 
exhibits amplification in the acceleration difference magnitude relative to the 
DNS signal, it retains the extreme, intermittent events that are largely 
smoothed by the discrete MLE approach.

The differences in these results can be attributed to the assumptions on the 
jerk distribution embedded in the formulation of each filter. The discrete MLE 
filter assumes a Gaussian jerk, and therefore penalizes large temporal 
variations in acceleration, leading to reduced acceleration differences. The 
IRLS formulation allows for sparse, high-magnitude changes in acceleration, 
resulting in the retention of the intermittent dynamics characteristic of 
turbulent flows.

\begin{figure}[t]
	\centering
	\includegraphics{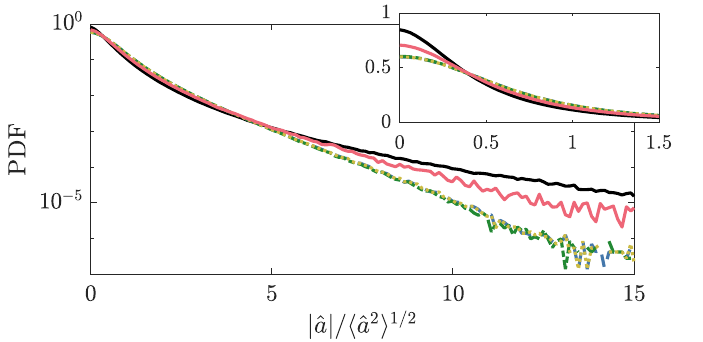}
	\caption{PDF of normalized accelerations. The DNS data exhibit the 
	characteristic heavy-tailed structure of turbulent acceleration. The 
	baseline MLE methods and the B-spline approach attenuate the tails of the 
	distribution, underpredicting the probability of extreme acceleration 
	events. In contrast, the IRLS reconstruction more closely follows the DNS 
	distribution, demonstrating improved recovery of the heavy-tail behavior. 
	The inset shows a zoomed view of the core of the distribution near zero 
	acceleration, highlighting differences in the central region.  Coloring and 
	line styles match those used in Figure~\ref{fig:pos}.}
	\label{fig:PDFa}
\end{figure}
\begin{figure}[h!]
	\centering
	\includegraphics{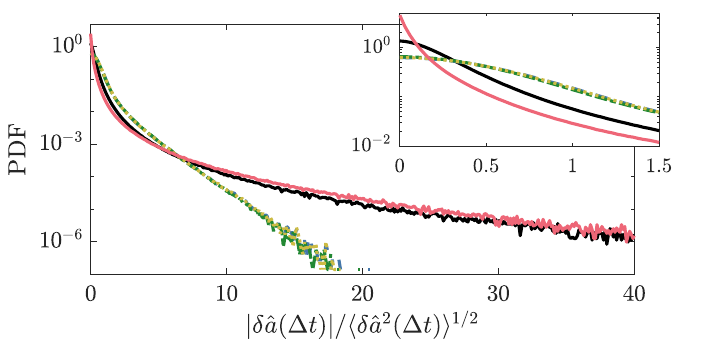}
	\caption{PDF of normalized acceleration differences evaluated at the 
	sampling interval. The DNS distribution exhibits pronounced heavy tails. 
	The baseline MLE and B-spline methods substantially compress the tails, 
	underpredicting the probability of high-magnitude acceleration increments. 
	In contrast, the IRLS reconstruction preserves a heavy-tailed structure 
	that closely follows the DNS statistics, particularly in the extreme-event 
	regime. The inset shows a zoomed view of the core of the distribution near 
	zero acceleration differences. Coloring and line styles are consistent with 
	Figure~\ref{fig:pos}.}
	\label{fig:PDFj}
\end{figure}

\begin{figure}[t]
	\centering
	\includegraphics{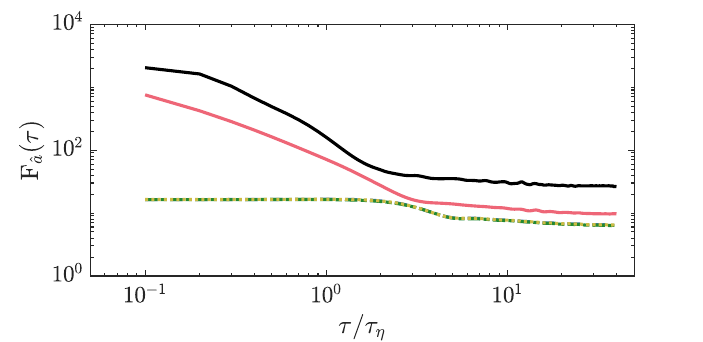}
	\caption{Flatness of normalized acceleration differences as a function of 
	separation time. The DNS data exhibit elevated flatness values at small 
	separations, reflecting strong intermittency and enhanced fourth-order 
	statistics. The baseline MLE and B-spline methods substantially reduce 
	flatness across scales, indicating suppression of high-magnitude 
	acceleration transitions. The IRLS reconstruction maintains significantly 
	higher flatness values that more closely follow the DNS behavior, 
	demonstrating improved preservation of intermittent dynamics across scales. 
	Coloring and line styles are consistent with Figure~\ref{fig:pos}.}
	\label{fig:Fa}
\end{figure}

\subsection{Acceleration Statistics}
\label{ssec:accel-dist}

Figure~\ref{fig:PDFa} shows the PDF of the normalized acceleration magnitude. 
The DNS distribution exhibits the well-known, heavy-tailed behavior, consistent 
with intermittent Lagrangian accelerations in turbulence. The discrete MLE, 
continuous MLE, and B-splines filtering approaches heavily attenuate the tails 
of the distribution, producing noticeably steeper decay in high volatility 
events. In contrast, the IRLS reconstruction closely follows the DNS 
distribution and substantially reduces attenuation of the tails, preserving the 
elevated probability of extreme events.

The distinction between filtering methods becomes most pronounced when 
examining the PDF of the acceleration differences evaluated at the sampling 
interval, shown in Figure~\ref{fig:PDFj}. This distribution reflects rapid 
temporal changes in acceleration and serves as a discrete proxy for jerk. The 
baseline filters dramatically compress the tails of this distribution, 
underpredicting extreme-events by several orders of magnitude. This compression 
indicates that rapid transitions are over-smoothed, fundamentally altering the 
measured small-scale dynamics. The IRLS filter preserves a heavy-tailed 
distribution, recovering the probability of large acceleration increments to a 
far greater extent compared to the baseline filters. 

Differences are observed in the core of the acceleration difference 
distribution, shown in the inset of Figure~\ref{fig:PDFj}. The IRLS filter 
exhibits an elevated probability near zero, reflecting a tendency to produce 
intervals of nearly constant acceleration, punctuated by intermittent jumps. By 
comparison, the baseline filters slightly underpredict the probability of 
near-zero increments, but maintain the Gaussian-like shape of the core. The 
$\ell_1$ term in the IRLS objective is dominant in the small jerk regime, and 
it mimics a Laplace distribution as is observed in this result. This 
characteristic causes the IRLS PDF to deviate from the Gaussian-like shape of 
the truth distribution for small jerk values. The behavior highlights an 
inherent trade-off in trajectory reconstruction: the IRLS filter model better 
predicts higher volatility events and better matches the true physics of the 
turbulent flow, but relaxation induces distribution deviations in the near zero 
jerk regime.
To further quantify intermittency across scales, we examine the flatness of 
acceleration differences as a function of separation time, shown in 
Figure~\ref{fig:Fa}. The flatness is given by $\mathrm{F}_{\hat{a}}(\tau) = 
\langle\delta \hat{a}^4(\tau)\rangle/\langle\delta \hat{a}^2(\tau)\rangle ^2$. 
High flatness values indicate strong intermittent behavior. Across all scales, 
baseline filtering methods exhibit markedly reduced flatness, consistent with 
suppression of higher-order structure. Again, the IRLS reconstruction maintains 
significantly larger flatness values that more closely track the DNS behavior, 
demonstrating that the improved tail recovery observed in the PDF is not 
limited to a single scale, but persists across temporal separations.

Taken together, these results show that baseline smoothing approaches mute 
intermittent dynamics, whereas our newly developed IRLS filter preserves the 
heavy-tailed statistical structure intrinsic to turbulent flows, while 
maintaining low global reconstruction error.

%%==================================%%
%%    CONCLUSIONS & FUTURE WORK     %%
%%==================================%%
\section{Conclusions and Future Work}
\label{sec:conclusion}

In this work, we have presented a modified MLE framework for Lagrangian 
particle tracking that explicitly accounts for intermittency in turbulent 
flows. By replacing standard Gaussian process noise assumptions with a 
heavy-tailed distribution modeled via a convex $\ell_1$-relaxation, we 
demonstrated a significant improvement in the recovery of flow physics. 
Specifically, the proposed IRLS filter successfully reconstructs the heavy 
tails of the acceleration PDF, which are systematically suppressed by 
conventional Gaussian smoothing and B-spline techniques, while maintaining 
position error strictly within the measurement noise floor. The results 
highlight improved performance in high-Reynolds-number flows when filtering 
strategies that are informed by richer underlying stochastic structures related 
to the turbulence.

Current results open several avenues for future research, particularly in 
establishing a rigorous link between the optimization hyperparameters and 
physical flow constants. While the sparsity penalty $\gamma$ and process noise 
$\sigma_v$ were tuned empirically in this study, they act as proxies for the 
physical intermittency and dissipation rate of the flow. Future experimental 
campaigns are warranted to characterize these parameters as functions of the 
Reynolds number ($Re_\lambda$) and Stokes number ($St$). We aim to develop a 
mapping that eliminates the need for post-hoc tuning, allowing the filter to 
adapt dynamically based on known flow conditions. Such a study would require an 
interdisciplinary approach, combining high-fidelity measurements from 
experimental facilities with advanced stochastic system identification.

From a theoretical perspective, the underlying dynamical model can be enriched 
beyond the third-order dynamics used here. As noted in the discussion, the 
current model assumes white process noise. However, the forces acting on a 
fluid particle often exhibit temporal correlations. Future iterations of this 
framework could incorporate colored noise models or additional physical 
constraints, such as damping terms or mean-field coupling, to better represent 
the particle's interaction with the viscous fluid. Integrating these 
higher-fidelity dynamical systems into the convex optimization framework would 
provide a more robust connection between the mathematical solver and the 
Navier-Stokes physics.

Finally, this framework provides a natural foundation for real-time state 
estimation in autonomous systems. While the IRLS algorithm presented here is 
iterative and suited for post-processing, it serves as a prototype for modern 
``algorithm unrolling'' techniques in machine learning. By unrolling the 
iterations of the IRLS solver into a deep neural network, it is possible to 
generate fixed-complexity, differentiable estimators that approximate the 
solution of the convex problem with millisecond latency. This translation from 
offline optimization to online inference has far-reaching consequences for flow 
control, enabling autonomous vehicles to estimate local flow structures and 
intermittent forcing events in real-time.
	
	\begin{appendices}
		\section{Error Landscapes in Filter Parameter Space}
		\label{sec:errorMap}
		
		\begin{figure}[h!]
			\centering
			\includegraphics{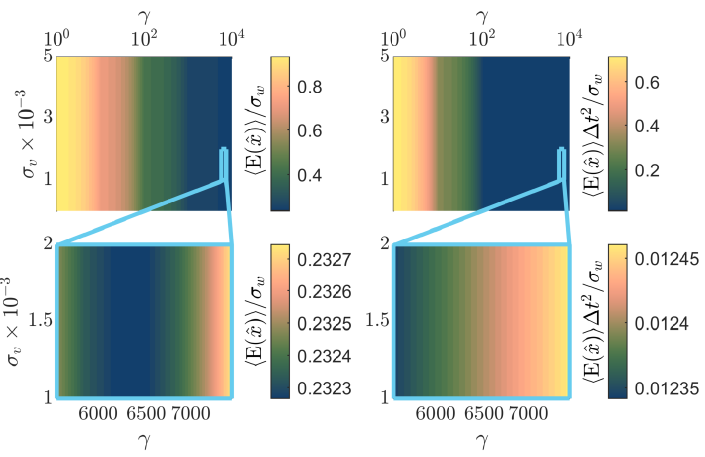}
			\caption{Hierarchical parameter tuning landscapes for position 
			(left) and acceleration (right). The coarse parameter sweep (top) 
			reveals that filter performance is highly sensitive to the sparsity 
			parameter $\gamma$ (log scale) while remaining relatively robust to 
			changes in $\sigma_v$ (vertical stratification). The 
			fine-resolution search (bottom) focuses on minor variations, 
			identifying an appealing performance configuration of $\gamma 
			\approx 5529$. The persistence of the vertical banding in the fine 
			stage confirms the filter's robustness to process noise variance 
			enabling a configuration with a relatively small $\sigma_v$ values.}
			\label{fig:tuning_combined}
		\end{figure}
		
		Figure \ref{fig:tuning_combined} shows the RMSE landscapes in the 
		$(\gamma,\sigma_v)$ parameter space for both position and acceleration. 
		The upper panels show the results of the initial coarse grid search, 
		which reveal a distinct sensitivity hierarchy. Error gradients are 
		primarily aligned with the $\gamma$ axis, resulting in vertical 
		stratification of the error surface. This indicates that reconstruction 
		quality is predominantly driven by the sparsity enforcement ($\gamma$) 
		rather than the specific process noise width ($\sigma_v$).
		
		The lower panels in Figure~\ref{fig:tuning_combined} present the 
		results from our fine-resolution search near the apparent minimum. 
		Although the optimal $\gamma$ values differ slightly between position 
		and acceleration, the error surface in this neighborhood is relatively 
		flat. Across the fine-resolution sweep, the performance varies by just 
		$3.7\times10^{-7}\%$ for position and $2.9\times10^{-8}\%$ for the 
		acceleration RMSE. The persistence of vertical banding at this scale 
		further reinforces the conclusions that the reconstruction performance 
		is relatively insensitive to $\sigma_v$ within the tested range.
		
		Taken together, these results indicate that the newly proposed IRLS 
		filter exhibits robust performance across a wide range of process noise 
		assumptions. This hierarchical sensitivity structure simplifies 
		practical parameter tuning in the absence of ground truth.
	\end{appendices}

	\bibliographystyle{plainnat} % or plainnat
	\bibliography{sn-bibliography}
\end{document}